\documentclass{article}      
\font\tensl=cmsy10
\usepackage{amssymb}
\title{\bf Quantization of Dirac fields in static spacetime}  
\author{Wei Min Jin \\ 
{\small Department of Physics and Astronomy, State University of New York at Buffalo,}\\
{\small Buffalo, NY 14260-1500, U.S.A.}\\
{\small Electronic mail: wjin1@excite.com}}  
\date{}      
\begin{document}             
\maketitle                   
\begin{center}
\begin{minipage}{135mm}
\vskip0.5in
\begin{center} {} \end{center}
{\bf Abstract}. On a static spacetime, the solutions of  
Dirac equation are generated by a time-independent Hamiltonian. 
We study this Hamiltonian and characterize the split 
into positive and negative energy.   We use it to find explicit 
expressions for advanced and retarded fundamental solutions and for the 
propagator. 
Finally we use a fermion Fock space based on the positive/negative energy 
split to define a Dirac quantum field operator whose commutator is the 
propagator. 

\end{minipage}
\end{center}
\vskip2in
\eject

\section* {1. Introduction}      

In the theoretical framework of
Dirac fields in curved space-time, many fundamental
results have been obtained by Lichnorowicz [1]. There are 
also quite a few standard references with detailed discussions, for
example, see [2][3][4]. A very general theory has been established by
Dimock [5], for Dirac quantum fields on hyperbolic Lorentzian
manifold. For most up-to-date research along this line,  
one can find recent papers by Hollands [6] and Kratzert [7], and
references therein. 

Here in this paper we focus on solving a very specific and well-posed problem:
the quantization of Dirac fields in static space-time. Technically speaking,
the separation of time and space parts of Dirac fields plays a pivotal
role in doing field quantization. Since only static space-time is considered
here, the following results about other type of fields are also of
relevance:  the quantization of Klein-Gordon scalar fields on
stationary manifold by Kay [8], and the quantization of electromagnetic
fields and massive vector fields on static space-time by Furlani
[9][10].  

This paper proceeds as follows. First we present some preliminary 
results for classical Dirac fields. 
In static space-time,  the time component of the spin 
affine connection of Dirac fields vanishes and the other components are all 
independent of time.  Hence one can separate space and time. The dynamics 
can be expressed in terms of a time-independent Hamiltonian. 
We prove that the Hamiltonian is essentially 
self-adjoint. We also characterize the positive and negative energy 
subspaces. This leads to explicit expressions for 
various fundamental solutions and for the propagator function. 

For the quantum problem,  we first 
define an appropriate fermion Fock 
space based on the positive/negative energy split. 
Then Dirac quantum field operators are defined using 
the creation and annihilation operators on this Fock 
space and projections onto the positive and negative energy 
subspaces. The field operator is shown to have 
a commutator which is the propagator function. 
Finally we study the unitary implementability of the 
time evolution.

\section* {2. Dirac fields in static space-time}

In this section, we present some preliminary results that are
relevant to our work. On Lorentzian manifold $L$, the first-order equation of free Dirac fields can be written as
$$(\hbox{\rlap/P}-m)\psi=0,\eqno(2.1a)$$
$$\psi^*(\hbox{\rlap/P}^*-m)=0,\eqno(2.1b)$$
where, by the notation in [11][12], $\psi^*=\psi^\dagger\gamma^0$ and $\hbox{\rlap/P}^*=\gamma^0\hbox{\rlap/P}^\dagger\gamma^0$ and
$$\hbox{\rlap/P}=i\gamma^\mu\nabla_\mu=i\gamma^\mu(\partial_\mu-\Gamma_\mu).\eqno(2.2)$$ 
Here $\gamma^\mu=V^\mu_a(x)\gamma^a$ are spinor tensors, with the introduction of vierbein
fields $V^\mu_a(x)$ and Dirac matrices $\gamma^a(a=0,1,2,3)$ by the
convention in [13]. The components of spin affine connection are
$$\Gamma_\mu={1\over2}G^{[a,b]}(\hbox{\tensl D}_\mu V^\nu_a)V_{\nu
b},\eqno(2.3)$$ 
where $G^{[a,b]}={1\over4}[\gamma^a,\gamma^b]$ are the generators of Lorentz
group, and $\hbox{\tensl D}_\mu V^\nu_a=\partial_\mu
V^\nu_a+\Gamma^\nu_{\mu\lambda}V^\lambda_a$ are the covariant derivatives of 
 vierbein fields on space.

The second-order equation of Dirac fields can be written as 
$$(\square -m^2)\psi=0,\eqno(2.4)$$
with an operator 
$$\square =\hbox{\rlap/P}^2=-\nabla^\mu\nabla_\mu+{1\over4}R,\eqno(2.5)$$
where $R$ is the Riemann scalar.

To preserve manifest covariance on Lorentzian Manifold $L$, we may introduce an
indefinite inner product [1]: 
$$<u,v>=\int u^*(x)v(x)d_g^4x,\hskip0.1in u,v\in C^\infty(L;C^4),\eqno(2.6)$$
where $d_g^4x=\sqrt{g}d^4x$ is the invariant density with
$g=-\det(g^{\mu\nu})$ and $u^*=u^\dagger\gamma^0$ is the adjoint of
$u$. The inner product is invariant under both global coordinate and local
Lorentz transformations. 

The adjoint operator $A^*$ of $A$ is defined by  
$$<u,A^*v>=<Au,v>,\eqno(2.7)$$
namely  
$$A^*=\gamma^0A^\dagger\gamma^0.\eqno(2.8)$$
A symmetric operator $O$ satisfies 
$$<u,Ov>=<Ou,v>,\eqno(2.9)$$
namely
$$O^*=\gamma^0O^\dagger\gamma^0=O.\eqno(2.10)$$
Symmetric operators play essential roles in functional analysis. 

Let us consider a static manifold $R\times M$ where $M$ is compact,
with metric elements 
$g^{\mu\nu}$ of signature $(1,-1,-1,-1)$,
$$[g^{\mu\nu}]=\left[\begin{array}{cc}1&0\\
0&g^{ij}({\bf x})\end{array}\right],\hskip0.1in(i,j=1,2,3)\eqno(2.11)$$ 
where Greek indices apply to 4-d static space-time, and Latin indices apply
to 3-d static space. The 3-d static space $M$ is a Cauchy surface of the 4-d
static space-time $R\times M$.

Let $u$ and $v$ have compact support. Then check Green's identity in static space-time $L=R\times M$ where
$M$ 
is compact without boundary 
$$<\hbox{\rlap/P}u,v>-<u,\hbox{\rlap/P}v>=\int_Ld_g^4x[(\hbox{\rlap/P}u)^*v-u^*\hbox{\rlap/P}v]$$
$$=-i\int_Ld_g^4x[(\nabla_\nu u)^*\gamma^\mu v+u^*\gamma^\mu\nabla_\mu v]$$
$$=-i\int_Ld_g^4x\nabla_\mu(u^*\gamma^\mu v),\eqno(2.12)$$
where $\nabla_\mu\gamma^\nu=0$ has been used. This integral is the
same as the integral over $[-T,T]\times M$ for $T$ sufficiently large
depending on the test functions. The integral of the divergence is an
integral over the boundary of this region which is $-T\times M$ and 
$T\times M$. The surface integrals are zero. We therefore have 
$$<\hbox{\rlap/P}u,v>=<u,\hbox{\rlap/P}v>,\eqno(2.13)$$
namely the operator {\rlap/P} is symmetric
$$\hbox{\rlap/P}^*=\hbox{\rlap/P}.\eqno(2.14)$$
Obviously $\Delta=\hbox{\rlap/P}^2$ is also symmetric
by the same arguments as for {\rlap/P}. 

The vierbein fields satisfy  
$g_{\mu\nu}(x)V^\mu_a(x)V^\nu_b(x)=\eta_{ab}$
with a Minkowski metric $\left\{\eta_{ab}\right\}=diag(1,-1,-1,-1)$.
 In static space-time, the vierbein
fields $V^0_0=1,\hskip0.05in V^0_i=0=V^i_0$ and $V^i_a({\bf
x})\hskip0.05in(i,a=1,2,3)$ are all independent of time. Then by (2.3)
it is easy to show that the time component of the spin affine
connection vanishes:  
$$\Gamma^\nu_{0\lambda}={1\over2}g^{\nu\sigma}(\partial_\lambda
g_{\sigma0}+\partial_0g_{\sigma\lambda}-\partial_\sigma g_{0\lambda})=0,$$
$$\hbox{\tensl
D}_0V^\nu_a=\partial_0V^\nu_a+\Gamma^\nu_{0\lambda}V^\lambda_a=0,$$ 
$$\Gamma_0={1\over2}G^{[a,b]}(\hbox{\tensl D}_0V^\nu_a)V_{\nu b}=0,\eqno(2.15)$$
and also the other components $\Gamma_i({\bf
x})\hskip0.05in(i=1,2,3)$ are all independent of time. 

Separating time from space, we assume there exists a global dreibein field
on $M$. If there is not a dreibein field, our analysis should still hold true,
however the spinor fields will be sections of a vector bundle [5].  
The simplest example of a compact manifold $M$ with a global dreibein field
is $M=T^3$, 
the three torus or periodic box. 

\section* {3. Separation of energy spectrum}

With the above results in static space-time, the Dirac equation
(2.1a) turns out to be  
$$i\partial_t\psi(t,{\bf x})=H\psi(t,{\bf x}),\eqno(3.1)$$
where a time-independent Hamiltonian is 
$$H=-i\gamma^0\gamma^i\nabla_i+\gamma^0m.\eqno(3.2)$$
And the Hamiltonian squared is
$$H^2=\nabla^i\nabla_i-{1\over4}R+m^2.\eqno(3.3)$$

Since we know $\gamma^i({\bf x})=V^i_a({\bf x})\gamma^a$ and
$-\gamma^0\gamma^a=\sigma^1\otimes\sigma^a$ where 
$\sigma^a$ are Pauli matrices. The Hamiltonian can then be
written 
in a matrix form 
$$H=\left[\begin{array}{cc}m&Q\\
Q&-m\end{array}\right],\eqno(3.4)$$
where we denote $Q=i\sigma^i\nabla_i$ and $\sigma^i=V^i_a({\bf
x})\sigma^a\hskip0.05in(i,a=1,2,3)$. 
The Hamiltonian squared becomes
$$H^2=(m^2+Q^2)\left[\begin{array}{cc}I&0\\
0&I\end{array}\right],\eqno(3.5)$$
where $Q^2=\nabla^i\nabla_i-{1\over4}R$.

Define a positive-definite inner product on $M$
$$(\chi,\varphi)=\int d_g^3{\bf x}(\chi^\dagger\varphi),\eqno(3.6)$$
where $d_g^3{\bf x}=\sqrt{g}d^3{\bf x}$. Then define a Hilbert space
$$\hbox{\tensl H}=L^2(M,C^4,d_g^3{\bf x}).\eqno(3.7)$$

Since the vierbein fields $V^i_a({\bf x})$ are
real functions of space on $M$ and Pauli matrices $\sigma^a$ are
symmetric on  
$C^2$, $\sigma^i=V^i_a({\bf x})\sigma^a$ are thus symmetric in
$L^2(M,C^2)$. It can be easily checked that ${G^{[a,b]}}^\dagger=-G^{[a,b]}$.
By (2.3), we know $(i\Gamma_i)^\dagger=i\Gamma_i$ and $i\Gamma_i$ are
some real functions
of space on $M$ after summation of indices. So
$i\nabla_i=i(\partial_i-\Gamma_i)$ is symmetric in $L^2(M)$, and 
$i\sigma^i\nabla_i$ is symmetric in $L^2(M,C^2)$. Then we know the
Hamiltonian $H$ is symmetric in $L^2(M,C^4)$.

Let $A$ be the closure of $H$ on $C^\infty(M,C^4)$ in $L^2(M,C^4)$. 
Define the domain of $A$ as follows
$$D(A)=\Bigl\{\psi\in\hbox{\tensl H}:\hskip0.05in\exists\psi_n\in C^\infty M,\hskip0.05in\lim_{n\rightarrow\infty}\psi_n\rightarrow\psi,\hskip0.05in
\lim_{n\rightarrow\infty}A\psi_n\hskip0.05in exists\Bigr\}.\eqno(3.8)$$
Then $A: D(A)\rightarrow\hbox{\tensl H}$ is given by defining
$$A\psi=\lim_{n\rightarrow\infty}A\psi_n.\eqno(3.9)$$ 
Let $B=A^2$ be the closure of $H^2$ on $C^\infty(M,C^4)$. 

\vskip0.1in
\noindent
{\it Lemma 1}. $B$ is self-adjoint in {\tensl H}, i.e. $H^2$ is essentially self-adjoint on $C^\infty(M,C^4)$.

\vskip0.1in
\noindent
{\it Proof}: It has been proven that the Laplacian operator $-\nabla^i\nabla_i$ is self-adjoint in
{\tensl H} [14][15]. Since $-{1\over4}R+m^2$ is a continuous function on a compact
manifold, it is a bounded function and hence a bounded operator in {\tensl H}
[16][17]. By Kato-Rellich theorem [18], $B$, a
closure of $H^2=\nabla^i\nabla_i-{1\over4}R+m^2$, is self-adjoint in {\tensl
H}. This is equivalent to say $H^2$ is essentially self-adjoint on 
$C^\infty(M,C^4)$. $\square$

\vskip0.1in
\noindent
{\it Lemma 2}. $A$ is self-adjoint in {\tensl H}, i.e. $H$ is essentially self-adjoint on $C^\infty(M,C^4)$.
\vskip0.1in
\noindent
{\it Proof}: This is equivalent to show [16]
$$Ran(A\pm i)=\hbox{\tensl H}. \eqno(3.10)$$
To prove it, we need to find $\psi$ so that 
$$(A\pm i)\psi=\chi,\hskip0.1in for\hskip0.1in \chi\in\hbox{\tensl
H}.\eqno(3.11)$$
By observation, the answer should be  
$$\psi=(A\mp i)(B+1)^{-1}\chi.\eqno(3.12)$$
It then suffices to show 
$$\varphi=(B+1)^{-1}\chi\in D(A),\eqno(3.13a)$$
$$\psi=(A\mp i)\varphi\in D(A).\eqno(3.13b)$$

By {\it Lemma 1}, for any
$\varphi\in D(B)$ we can find smooth $\varphi_n$ so that
$$\lim_{n\rightarrow\infty}\varphi_n\rightarrow\varphi,\hskip0.1in\lim_{n\rightarrow\infty}B\varphi_n\rightarrow
B\varphi.\eqno(3.14)$$
Now we derive  
$$\parallel A\varphi_n-A\varphi_m\parallel^2=(\varphi_n-\varphi_m,A^2(\varphi_n-\varphi_m))$$
$$\leq\parallel\varphi_n-\varphi_m\parallel\times\parallel
B(\varphi_n-\varphi_m)\parallel\rightarrow0.\eqno(3.15)$$
So we know $\varphi\in D(A)$ by (3.8). Then we may define smooth 
$\psi_n=(A\mp i)\varphi_n$ so that $\psi_n \rightarrow\psi$ and derive
$$\parallel A\psi_n-A\psi_m\parallel^2=(A^2(\varphi_n-\varphi_m),(A^2+1)(\varphi_n-\varphi_m))$$
$$\leq\parallel B(\varphi_n-\varphi_m)\parallel\times\parallel
(B+1)(\varphi_n-\varphi_m)\parallel\rightarrow0.\eqno(3.16)$$
Thus we know $\psi\in D(A)$. By the above analysis, $A$ is self-adjoint
in {\tensl H}, and equivalently $H$ is essentially self-adjoint on 
$C^\infty(M,C^4)$. $\square$

\vskip0.1in
By {\it Lemma 1}, $H^2$ is
essentially self-adjoint and positive, and hence has a square root. We may
 define a positive scalar energy operator
$$\omega=(m^2+Q^2)^{1/2}=(\nabla^i\nabla_i-{1\over4}R+m^2)^{1/2},\eqno(3.17)$$
where $Q=i\sigma^i\nabla_i$ for $\sigma^i=V^i_a({\bf x})\sigma^a$.
By {\it Lemma 2}, the closure
of $H$ is self-adjoint, and similarly the closure of $Q$ is
self-adjoint, then $Q^2=\nabla^i\nabla_i-R/4\geq0$ and $\omega\geq m$.
It is now straightforward to prove the following Theorem.

\vskip0.1in
\noindent
{\it Theorem 1}: The Hilbert space {\tensl H} splits into the positive and negative
subspaces: $\hbox{\tensl H}=\hbox{\tensl H}^+\oplus\hbox{\tensl H}^-$. Then
for $\psi_\pm\in\hbox{\tensl H}^\pm\cap D(H)$, we have
$$H\psi_\pm=\pm\omega\psi_\pm,\eqno(3.18)$$
where $\omega$ is expressed by (3.17). And also $\psi_\pm$ are of the
following form: 
$$\psi_+=T\left[\begin{array}{c}f\\
0\end{array}\right],\hskip0.1in\psi_-=T\left[\begin{array}{c}0\\
h\end{array}\right],\hskip0.1in f,h\in L^2(M;C^2),\eqno(3.19)$$
where $T$ is the unitary operator
$$T=N\left[\begin{array}{cc}{\omega+m}&-Q\\
Q&{\omega+m}\end{array}\right],\eqno(3.20)$$
with $N=[2\omega(\omega+m)]^{-1/2}$.

\vskip0.1in
\noindent
{\it Proof}: We diagonalize the Hamiltonian (3.4) to 
$$H'=T^{-1}HT=\left[\begin{array}{cc}\omega&0\\
0&-\omega\end{array}\right],\eqno(3.21)$$
by using a transformation operator (3.20). The inverse of
$T$ is  
$$T^{-1}=N\left[\begin{array}{cc}{\omega+m}&Q\\
-Q&{\omega+m}\end{array}\right].\eqno(3.22)$$
Both $T$ and $T^{-1}$ are norm-preserving: 
$$\int(T\varphi)^\dagger T\varphi d\mu=\int\varphi^\dagger\varphi d\mu=\int(T^{-1}\varphi)^\dagger T^{-1}\varphi d\mu, \eqno(3.23)$$ 
so $T$ is unitary. 

Any $\psi$ in {\tensl H} can be written as 
$$\psi=T\left[\begin{array}{c}f\\
h\end{array}\right],\hskip0.1in f,h\in L^2(M;C^2)\eqno(3.24)$$ 
and hence $\psi=\psi_+ +\psi_-$ where
$$\psi_+=T\left[\begin{array}{c}f\\
0\end{array}\right],\hskip0.1in \psi_-=T\left[\begin{array}{c}0\\
h\end{array}\right],\hskip0.1in f,h\in L^2(M;C^2).\eqno(3.25)$$
If also $\psi_\pm \in D(H)$, then
$$H\psi_\pm=\pm\omega\psi_\pm.\eqno(3.26)$$
Thus we have exhibited the split into the positive and negative energy
spectra. (Note 
that 0 is not in the spectra since $\omega\geq m>0$.) $\square$

\vskip0.1in
By the above {\it Theorem 1}, we can now express Dirac fields by summing up both
positive and negative energy parts
$$\psi(t,{\bf x})=U(t)\psi_+({\bf x})+U(-t)\psi_-({\bf x}),\eqno(3.27)$$
which is a solution of Dirac equation with data $\psi$. Here $\psi_\pm$
satisfy (3.25), and a unitary operator  
$U(t)=\exp(-i\omega t)$ satisfies $U^\dagger(t)=U(-t)=U^{-1}(t)$ with $\omega$ given by (3.17). 

\section* {4. Propagator of Dirac equation}

To obtain the propagator of Dirac equation, we start from the second-order inhomogeneous equation 
$$(\square-m^2)\psi=\rho,\eqno(4.1)$$
where $\square$ is expressed by (2.5). The fundamental solutions
of this inhomogeneous equation are defined by 
$$(\square_x-m^2)E_F(x,y)=\delta(x,y).\eqno(4.2)$$
Here the $\delta$-function is defined by a bispinor [1] 
$$\delta(x,y)=\delta^\alpha_\beta\delta^4(x,y).\eqno(4.3)$$

The general discussions about the existence and uniqueness of
fundamental solutions of hyperbolic differential equations such as (4.1) can be found in
[3][4]. Now we are going to do a formal calculation to find a
representation for the advanced and retarded fundamental solutions $E_A, E_R$.

On static metric $R\times M$, we separate
the time and space parts of the solutions.  It is
translation-invariant along time direction but not necessarily along space
direction. We make Fourier transform along time axis but leave 
the space part alone.  
The fundamental solutions can be written as 
$$E_F(x,y)={1\over2\pi}\int e^{-ikt}E_k({\bf x},{\bf y})dk,\eqno(4.4)$$
where
$$t=x_0-y_0. \eqno(4.5)$$
Inserting (4.4) into (4.2) leads to 
$$(k^2-\nabla^i\nabla_i+{1\over4}R-m^2)E_k({\bf x},{\bf
y})=\delta({\bf x},{\bf y}).\eqno(4.6)$$
We formally write k-component solution 
$$E_k({\bf x},{\bf y})=(k^2-\omega^2)^{-1}\delta({\bf x},{\bf y}),\eqno(4.7)$$
with $\omega^2=\nabla^i\nabla_i-R/4+m^2$. So (4.4) formally becomes
$$E_F(x,y)={1\over2\pi}\int{e^{-ikt}\over{k^2-\omega^2}}\delta({\bf
x},{\bf y})dk,\eqno(4.8)$$
which is singular with a delta function. Let us smear it by two
test functions 
$\chi({\bf x})$ and $\varphi({\bf y})$ on $M$
$$E_F(t;\chi,\varphi)={1\over2\pi}\int
e^{-ikt}(\chi,{1\over{k^2-\omega^2}}\varphi)dk.\eqno(4.9)$$
Let $P_\lambda=P_{(-\infty,\lambda]}$ be a projection-valued measure of self-adjoint
 operator $\omega$. Its family $\left\{P_\lambda\right\}$ exists by the
 Spectral Theorem [16]. Thus (4.9) becomes\footnote{Since we are
 considering compact manifolds, the spectral integral is actually a
 sum over discrete eigenvalues. Nevertheless we stick with the more
 general notation.}   
$$E_F(t;\chi,\varphi)={1\over2\pi}\int\int{e^{-ikt}\over{k^2-\lambda^2}}dk(\chi,dP_\lambda
\varphi),\hskip0.1in \forall\chi,\varphi\in\hbox{\tensl H}.\eqno(4.10)$$
We may now prove the following proposition.

\vskip0.1in
\noindent
{\it Proposition 1}: The advanced and retarded
fundamental solutions in integral representation, defined by 
$$E_A(t;\chi,\varphi)={1\over2\pi}\int\int_{\Gamma_A}{e^{-ikt}\over{k^2-\lambda^2}}dk(\chi,dP_\lambda
\varphi),\hskip0.1in \forall\chi,\varphi\in\hbox{\tensl H},\eqno(4.11a)$$
$$E_R(t;\chi,\varphi)={1\over2\pi}\int\int_{\Gamma_R}{e^{-ikt}\over{k^2-\lambda^2}}dk(\chi,dP_\lambda
\varphi),\hskip0.1in \forall\chi,\varphi\in\hbox{\tensl H},\eqno(4.11b)$$
vanish in the future and past respectively.
Here $\Gamma_A$ ($\Gamma_R$) is a straight line slightly
below (above) the real k-axis.   

\vskip0.1in
\noindent
{\it Proof}: Obviously (4.10) has two poles $k=\pm\lambda$ for each $\lambda$
in the 
integral over $k$. Let us avoid two poles in (4.11a) by taking the integral along a
straight line $\Gamma_A$ which passes 
slightly below the real k-axis. This is equivalent to moving the poles to
slightly above the real k-axis. When $t>0$, we 
calculate the integral by closing the contour in the lower half of the
complex k-plane. Since $Re(-ikt)<0$, the infinite semicircle boundary in the
lower half plane does not contribute. And there is no pole inside the closed
contour, $E_A$ vanishes in the future when $t>0$. 
Similarly $E_R$ defined by (4.11b) vanishes in the past when $t<0$, by
 the integral along a straight line $\Gamma_R$ which passes slightly above the real k-axis. It is also
straightforward to check that the expressions are actually fundamental
solutions. From [3][4], the fundamental solutions of hyperbolic
differential equations have support only in light cone. So the advanced and
retarded fundamental solutions $E_A$ and $E_R$ have support only in
the past and future light cones respectively. $\square$

\vskip0.1in
Let us take the difference of (4.11a) and (4.11b) by closing contour $C$
around both poles 
$$E(t;\chi,\varphi)=E_A(t;\chi,\varphi)-E_R(t;\chi,\varphi)$$
$$={1\over2\pi}\int\oint_C{e^{-ikt}\over{k^2-\lambda^2}}dk(\chi,dP_\lambda
\varphi).\eqno(4.12)$$
It can also be split into the positive and negative parts by
choosing the contours around both poles: 
$$E(t;\chi,\varphi)=E_+(t;\chi,\varphi)+E_-(t;\chi,\varphi), \eqno(4.13)$$
where
$$E_\pm(t;\chi,\varphi)={1\over2\pi}\int\oint_{C_\pm}{e^{-ikt}\over{k^2-\lambda^2}}dk(\chi,dP_\lambda
\varphi)$$
$$=\pm i\int{e^{\mp i\lambda t}\over{2\lambda}}(\chi,dP_\lambda
\varphi)$$
$$=\pm i(\chi, {e^{\mp i\omega
t}\over{2\omega}}\varphi).\eqno(4.14)$$ 

The fundamental solutions of the inhomogeneous Dirac equation 
$$(\hbox{\rlap/P}-m)\psi=\rho,\eqno(4.15)$$
are defined by 
$$(\hbox{\rlap/P}-m)S_F(x,y)=\delta(x,y).\eqno(4.16)$$
From $(\hbox{\rlap/P}-m)(\hbox{\rlap/P}+m)=\square-m^2$, we see
$$S_F(x,y)=(\hbox{\rlap/P}+m)E_F(x,y),\eqno(4.17)$$ 
which can be smeared by two test functions
$\chi({\bf x})$ and $\varphi({\bf y})$ on $M$:
$$S_F(t;\chi,\varphi)={1\over2\pi}\int
e^{-ikt}dk(\chi,{{\hbox{\rlap/P}(k)+m}\over{k^2-\omega^2}}\varphi),\eqno(4.18)$$
where the operator {\rlap/P}$(k)$ is also a function of $k$:
$$\hbox{\rlap/P}(k)=\gamma^0k+i\gamma^i\nabla_i.\eqno(4.19)$$

Following the discussions about $E_\pm$ (4.14), we
obtain 
$$S_\pm(t;\chi,\varphi)={1\over2\pi}\int\oint_{C_\pm}e^{-ikt}dk(\chi,{{\hbox{\rlap/P}(k)+m}\over{k^2-\lambda^2}}dP_\lambda
\varphi)$$
$$=\pm i\int e^{\mp i\lambda t}(\chi,{{\hbox{\rlap/P}(\pm\lambda)+m}\over{2\lambda}}dP_\lambda
\varphi)$$
$$=(\chi,ie^{\mp i\omega t}\pi_\pm\gamma^0\varphi).\eqno(4.20)$$ 
Here $\pi_\pm$ are two orthogonal projection operators
onto the positive and negative energy parts in Hilbert space respectively, 
$$\pi_\pm={{\omega\pm(-i\gamma^0\gamma^i\nabla_i+\gamma^0m)}\over2\omega}.\eqno(4.21)$$
We can check they are the correct projection operators by applying them on the positive
and negative energy solutions (3.19):
$$\pi_\pm\psi_\pm=\psi_\pm,\hskip0.1in\pi_\pm\psi_\mp=0.\eqno(4.22)$$
Generally, $\pi_\pm$ have the following relations:
$$\pi^\dagger_\pm=\pi_\pm=\pi^2_\pm,\eqno(4.23a)$$
$$\pi_\pm\pi_\mp=0,\eqno(4.23b)$$
$$\pi_++\pi_-=1.\eqno(4.23c)$$ 
And $\pi_\pm$ commute with $\omega$, 
since $\omega^2$ commutes with $\gamma^i\nabla_i$ and $\omega$ is a scalar.

By (4.20), the positive and negative energy parts of 
$S$-function in space-time representation can be formally expressed by
$$S_\pm(x,y)=ie^{\mp i\omega t}\pi_\pm({\bf x})\gamma^0\delta({\bf x},{\bf
y}),\eqno(4.24)$$
which make sense mathematically only if they are smeared as in (4.20).

Similar to $E(x,y)$, the propagator of Dirac equation is defined as the difference between the advanced and
retarded fundamental solutions:
$$S(x,y)=S_A(x,y)-S_R(x,y).\eqno(4.25)$$
where $S_A, S_R$ are related to $E_A, E_R$ by (4.17),
or the summation of the positive and negative energy fundamental solutions:
$$S(x,y)=S_+(x,y)+S_-(x,y).\eqno(4.26)$$
Since $supp(S_Ru)\cap
supp(S_Av)$ is compact, by using the same 
procedure of deriving equation (2.13) we compute 
$$<S_Ru,v>=<S_Ru,(\hbox{\rlap/P}-m)S_Av>$$
$$=<(\hbox{\rlap/P}-m)S_Ru,S_Av>=<u,S_Av>.\eqno(4.27)$$
Here we can see the advanced and retarded
fundamental solutions are the adjoints of each other in spin space
$$[S_R(x,y)]^*=S_A(y,x),\eqno(4.28)$$
and similarly
$$[S_\pm(x,y)]^*=-S_\pm(y,x).\eqno(4.29)$$
Therefore the propagator satisfies
$$[S(x,y)]^*=-S(y,x),\eqno(4.30)$$
and there is an obvious result 
$$[-iS(x,y)]^*=-iS(y,x).\eqno(4.31)$$
This completes our discussions on the propagator of Dirac equation.

\section* {5. Quantization of Dirac fields}

To construct field theory one
needs to define a Fock space with one particle space as its base space
[19]. A general Fock space is defined by [20][21]
$$\hbox{\tensl F}(\hbox{\tensl H})=\oplus_{n=0}^\infty\hbox{\tensl F}^{(n)}(\hbox{\tensl
H})=\hbox{\tensl F}^{(0)}(\hbox{\tensl H})\oplus\dots\oplus\hbox{\tensl
F}^{(n)}(\hbox{\tensl H})\oplus\dots,\eqno(5.1)$$
where n-fold tensor subspace is 
$$\hbox{\tensl F}^{(n)}(\hbox{\tensl H})=\otimes\hbox{\tensl
H}^{(n)}=\hbox{\tensl H}\otimes\dots\otimes\hbox{\tensl
H},\eqno(5.2)$$
and $\hbox{\tensl F}^{(0)}(\hbox{\tensl H})=C$ is the vacuum space with
complex constants as its elements. {\tensl H} is any complex Hilbert space
with a positive-definite inner product. In Fock space $\psi\in\hbox{\tensl
F}(\hbox{\tensl H})$
$$\psi=(\psi^{(0)},\psi^{(1)},\dots\psi^{(n)},\dots),\eqno(5.3)$$
where $\psi^{(n)}\in\hbox{\tensl F}^{(n)}(\hbox{\tensl H})$. A dense set in
$\hbox{\tensl F}^{(n)}(\hbox{\tensl H})$ is linear combinations of vectors of
the form
$$\psi^{(n)}=\psi_1\otimes\dots\otimes\psi_n.\eqno(5.4)$$
The inner product on $\hbox{\tensl F}(\hbox{\tensl H})$ is induced by the
inner product on $\hbox{\tensl H}$: 
$$(\psi,\psi)_{\hbox{\tensl
F}}=|\psi^{(0)}|^2+(\psi^{(1)},\psi^{(1)})_{\hbox{\tensl
H}^{(1)}}+\dots+(\psi^{(n)},\psi^{(n)})_{\hbox{\tensl H}^{(n)}}+\dots.\eqno(5.5)$$
If $\psi^{(n)}$ has the form (5.4), then
$$(\psi^{(n)},\psi^{(n)})_{\hbox{\tensl
H}^{(n)}}=\prod^n_{i=1}(\psi_i,\psi_i)_{\hbox{\tensl H}}.\eqno(5.6)$$

The construction of Fock space can be put in the above form for both
fermion 
and boson fields. Here we are only interested in Dirac fermion fields which
obey antisymmetric rule. Define linear permutation operators on {\tensl F}({\tensl H}) by 
$$\Pi(\psi_1\otimes\dots\otimes\psi_n)={1\over n!}\sum_\pi(-1)^\pi\psi_{\pi(1)}\otimes\dots\otimes\psi_{\pi(n)},\eqno(5.7)$$
which induces orthogonal projections onto $\hbox{\tensl F}^{(n)}$({\tensl
H}). For example, 
$$\Pi(\psi_1\otimes\psi_2)={1\over2}(\psi_1\otimes\psi_2-\psi_2\otimes\psi_1).\eqno(5.8)$$
Applying this operator to {\tensl F}({\tensl H}), we obtain 
antisymmetric fermionic Fock space 
$$\hbox{\tensl F}_a(\hbox{\tensl H})=\oplus_{n=0}^\infty\hbox{\tensl F}^{(n)}_a(\hbox{\tensl H}),\eqno(5.9)$$
in which n-fold subspaces are defined by 
$$\hbox{\tensl F}^{(n)}_a(\hbox{\tensl H})=\Pi\hbox{\tensl
F}^{(n)}(\hbox{\tensl H}).\eqno(5.10)$$

To have a complete description of Fock space, one should define creation
and annihilation operators. On the vacuum space $\hbox{\tensl
F}^{(0)}(\hbox{\tensl H})$, one defines 
$$a_0(\chi)\psi_0=0,\eqno(5.11a)$$
$$a_0^\dagger(\chi)\psi_0=\chi.\eqno(5.11b)$$
Generally, one can define the
creation and annihilation operators on
$\hbox{\tensl F}(\hbox{\tensl H})$ by
$$a_0^\dagger(\chi)(\psi_1\otimes\dots\otimes\psi_n)=\sqrt{n+1}(\chi\otimes\psi_1\otimes\dots\otimes\psi_n),\eqno(5.12a)$$
$$a_0(\chi)(\psi_1\otimes\dots\otimes\psi_n)=\sqrt{n}(\chi,\psi_1)_{\hbox{\tensl
H}}(\psi_2\otimes\dots\otimes\psi_n).\eqno(5.12b)$$
Then on Fermi-Fock space $\hbox{\tensl F}_a(\hbox{\tensl H})$, according to
Bratteli and Robinson [22], the
creation operator $a^\dagger(\chi)$ and annihilation operator $a(\chi)$ can
be defined as
$$a^\dagger(\chi)=\Pi a_0^\dagger(\chi)\Pi,\eqno(5.13a)$$
$$a(\chi)=\Pi a_0(\chi)\Pi.\eqno(5.13b)$$
It is straightforward to show the creation and annihilation operators
satisfy the following canonical anticommutation relations (CAR)
$$\Bigl\{a(\chi),a(\varphi)\Bigr\}=0,\eqno(5.14a)$$ 
$$\Bigl\{a^\dagger(\chi),a^\dagger(\varphi)\Bigr\}=0,\eqno(5.14b)$$
$$\Bigl\{a(\chi),a^\dagger(\varphi)\Bigr\}=(\chi,\varphi)_{\hbox{\tensl
H}},\eqno(5.14c)$$
where $(\chi,\varphi)_{\hbox{\tensl
H}}$ represents an inner product in Hilbert space.

Let us define a Fermi-Fock space in static space-time   
$$\hbox{\tensl F}_a(\hbox{\tensl H})=\hbox{\tensl F}_a(\hbox{\tensl H}_+)\otimes\hbox{\tensl F}_a(\hbox{\tensl H}_-),\eqno(5.15)$$
where $\hbox{\tensl F}_a(\hbox{\tensl H}_\pm)$ are the positive and negative
subspaces respectively. Also define 
$$a_+(\chi)=a(\chi)\otimes I, \hskip0.1in a^\dagger_+(\chi)=a^\dagger(\chi)\otimes
I;\eqno(5.16a)$$
$$a_-(\chi)=(-1)^N\otimes a(\chi), \hskip0.1in a^\dagger_-(\chi)=(-1)^N\otimes
a^\dagger(\chi), \eqno(5.16b)$$
where $(-1)^N$ is necessary so that $a_+$ and $a_-$ anti-commute [20], and
$a(\chi), a^\dagger(\chi)$ are defined in (5.13) with inner product
$(\chi,\varphi)_{\hbox{\tensl H}}=(\chi,\varphi)$. It is
straightforward to check that the so-defined 
annihilation and creation operators satisfy the following CAR 
$$\Bigl\{a_\pm(f_\pm),a_\pm(h_\pm)\Bigr\}=0,\eqno(5.17a)$$ 
$$\Bigl\{a_\pm^\dagger(f_\pm),a_\pm^\dagger(h_\pm)\Bigr\}=0,\eqno(5.17b)$$
$$\Bigl\{a_\pm(f_\pm),a_\pm^\dagger(h_\pm)\Bigr\}=(f_\pm,h_\pm),\eqno(5.17c)$$
where $f_\pm$ are any vectors in the positive and negative energy
subspaces respectively.

The classical solutions of the Dirac equation are given by (3.27). 
Correspondingly we define a quantum field operator to be the solution 
with data which are the positive and negative energy annihilation 
operators  $a_\pm({\bf x}), a_\pm^\dagger({\bf x})$ given formally by 
$$a_\pm(\chi)=\int \chi^\dagger({\bf x})a_\pm({\bf x})d_g^3{\bf 
x},\eqno(5.18a)$$ 
$$a_\pm^\dagger(\chi)=\int a_\pm^\dagger({\bf x})\chi({\bf x})d_g^3{\bf 
x},\eqno(5.18b)$$ 
where $d_g^3{\bf x}=\sqrt{g}d^3{\bf x}$.
Thus we put 
$$\psi(t,{\bf x})=U(t)\pi_+a_+({\bf x})+U(-t)\pi_-a_-({\bf 
x}),\eqno(5.19)$$ 
where  $U(\mp t)=\exp(\pm i\omega t)$.
Smearing $\psi$ by a $C^\infty_0$ test function $f(t,{\bf x})$ on 
$R\times M$,  we get the field
$$\psi(f)=\int f^*(x)\psi(x)d_g^4x,\eqno(5.20a)$$ 
and also by our convention $\psi^*=\psi^\dagger\gamma^0$,
$$\psi^*(f)=\int\psi^*(x)f(x)d_g^4x,\eqno(5.20b)$$
with a relation 
$$\psi^*(f)= \psi(f)^\dagger.\eqno(5.21)$$ 
Then by (5.19), it is easy to see (5.20) leads to
$$\psi(f)=a_+(f_+)+a_-(f_-),\eqno(5.22a)$$ 
$$\psi^*(f)=a^\dagger_+(f_+)+a^\dagger_-(f_-),\eqno(5.22b)$$
where $f_\pm$ on $M$ are the positive and negative energy components defined 
by   
$$f_\pm({\bf x})=\int e^{\pm i\omega x_0}\pi_\pm\gamma^0f(x_0,{\bf 
x})dx_0.\eqno(5.23)$$ 
We take (5.22) as the precise definitions of $\psi(f)$
and $\psi^*(f)$. With the above preparations, we are now ready to
prove the following theorem.

\vskip0.1in
\noindent
{\it Theorem 2}: In static space-time, let Dirac field operators be expressed
in terms of the 
creation and 
annihilation operators on Fermi-Fock space as in (5.22).
Then the quantized Dirac field operators satisfy the equation
$$\psi((\hbox{\rlap/P}-m)f)=0,\eqno(5.24a)$$
$$\psi^*((\hbox{\rlap/P}-m)h)=0,\eqno(5.24b)$$
and the following CAR
$$\Bigl\{\psi(f),\psi(h)\Bigr\}=0,\eqno(5.25a)$$
$$\Bigl\{\psi^*(f),\psi^*(h)\Bigr\}=0,\eqno(5.25b)$$
$$\Bigl\{\psi(f),\psi^*(h)\Bigr\}=-i<f,Sh>,\eqno(5.25c)$$
together with an integral
$$<f,Sh>=\int\int f^*(x)S(x,y)h(y)d_g^4xd_g^4y,\eqno(5.26)$$ 
in terms of a propagator $S(x,y)$ obtained in \S4.

\vskip0.1in
\noindent
{\it Proof}: Let $D=\hbox{\rlap/P}-m$. We can write 
$$\psi(Df)=a_+((Df)_+)+ a_-((Df)_-).\eqno(5.27)$$
By (5.23), it follows by integration by parts
$$(Df)_\pm({\bf x})=\int e^{\pm i\omega x_0}\pi_\pm\gamma^0D_\pm f(x_0,{\bf
x})dx_0,\eqno(5.28)$$
where $\pi_\pm$ are expressed in (4.21) and
$$D_\pm=\pm\gamma^0\omega+i\gamma^i\nabla_i-m=\pm2\omega\gamma^0\pi_\mp.\eqno(5.29)$$  
It is then clear to see
$$\pi_\pm\gamma^0D_\pm=\pm2\omega\pi_\pm\pi_\mp=0,\eqno(5.30)$$
So we end up with (5.24).

By (5.22) and (5.17a) and (5.17b), it is easy to show (5.25a) and (5.25b).
To show the nonvanishing CAR (5.25c), we first compute by using
(4.20), (4.23)
and (5.23) 
$$-i<f,S_\pm h>=-i\int\int S_\pm(x_0-y_0; \gamma^0f(x_0,\cdot),
h(y_0,\cdot))dx_0dy_0$$
$$=\int\int(\gamma^0f(x_0,\cdot),e^{\mp
  i\omega(x_0-y_0)}\pi_\pm\gamma^0h(y_0,\cdot))dx_0dy_0$$
$$=\int\int(e^{\pm i\omega x_0}\pi_\pm\gamma^0f(x_0,\cdot),e^{\pm i\omega y_0}\pi_\pm\gamma^0h(y_0,\cdot))dx_0dy_0$$
$$=(f_\pm,h_\pm).\eqno(5.31)$$
Then using (5.22), (5.17), (5.31) and (4.26), we get 
$$\Bigl\{\psi(f),\psi^*(h)\Bigr\}=\Bigl\{a_+(f_+),a_+^\dagger(h_+)\Bigr\}+\Bigl\{
a_-(f_-),a_-^\dagger(h_-)\Bigr\}$$
$$=(f_+,h_+)+(f_-,h_-)$$
$$=-i<f,S_+h>-i<f,S_-h>$$
$$=-i<f,Sh>,\eqno(5.32)$$
which is just (5.25c). $\square$

\vskip0.1in
If we take the sum
$$f_s=f_++f_-,\eqno(5.33)$$ 
then we obtain
$$-i<f,Sh>=(f_+,h_+)+(f_-,h_-)=(f_s,h_s).\eqno(5.34)$$
We may just take $\psi(f)$ and $\psi^*(h)$ as the annihilation and
creation operators on Fermi-Fock space $\hbox{\tensl F}_a(\hbox{\tensl H})$ and define
$$\Bigl\{\psi(f),\psi^*(h)\Bigr\}=(f_s,h_s).\eqno(5.35)$$
This becomes a special case of Dimock's general theorem [5].

Considering Dirac fields as operators, under time
evolution we would have by Heisenberg picture
$$\psi(x_0+t,{\bf x})=e^{iKt}\psi(x_0,{\bf x})e^{-iKt},\eqno(5.36a)$$
$$\psi^*(x_0+t,{\bf x})=e^{iKt}\psi^*(x_0,{\bf x})e^{-iKt},\eqno(5.36b)$$
where $K$ is the energy operator. Smearing them by test function $f(x_0,{\bf x})$ it is easy to see the left
sides of (5.36) become 
$$\int f^*(x_0-t,{\bf x})\psi(x_0,{\bf x})d_g^4x=\psi(f(\cdot -t)),\eqno(5.37a)$$
$$\int \psi^*(x_0,{\bf x})f(x_0-t,{\bf x})d_g^4x=\psi^*(f(\cdot -t)),\eqno(5.37b)$$
and the right sides of (5.36) become 
$$e^{iKt}[\int f^*(x_0,{\bf x})\psi(x_0,{\bf
x})d_g^4x]e^{-iKt}=e^{iKt}\psi(f)e^{-iKt},\eqno(5.38a)$$
$$e^{iKt}[\int \psi^*(x_0,{\bf x})f(x_0,{\bf
x})d_g^4x]e^{-iKt}=e^{iKt}\psi^*(f)e^{-iKt}.\eqno(5.38b)$$
So we should get
$$\psi_t(f)=\psi(f(\cdot -t))=e^{iKt}\psi(f)e^{-iKt},\eqno(5.39a)$$
$$\psi_t^*(f)=\psi^*(f(\cdot -t))=e^{iKt}\psi^*(f)e^{-iKt}.\eqno(5.39b)$$
Now we show how this works out.

Define unitary operators for time
evolution by
$$U(t)=U^+(t)\otimes U^-(t),\eqno(5.40)$$
and
$$U^\pm(t)=\oplus_{n=0}^\infty U_n^\pm(t),\eqno(5.41)$$
and
$$U_n^\pm(t)=e^{\mp i\omega t}\otimes\dots\otimes e^{\mp i\omega t}.\eqno(5.42)$$
These are all unitary groups with generators $K$, $K^\pm$ and $K_n^\pm$
respectively. Here $K$ is given by
$$K=K^+\otimes I+ I\otimes K^-,\eqno(5.43)$$
which is not positive-definite, and $K^\pm$ is given by
$$K^\pm=\oplus_{n=0}^\infty K^\pm_n,\eqno(5.44)$$
and $K^\pm_n$ is given by 
$$K^\pm_n=\omega\otimes I\otimes\dots\otimes
I+I\otimes\omega\otimes I\otimes\dots\otimes I$$
$$+\dots+I\otimes\dots\otimes I\otimes\omega\hskip0.2in(\hbox{n}\hskip0.05in \hbox{terms}).\eqno(5.45)$$
We now want to prove (5.39) in terms of the time evolution of creation
and annihilation operators by a proposition. 

\vskip0.1in
\noindent
{\it Proposition 2}: With the above definitions of the Dirac field operators
$\psi(f),\psi^*(f)$ and the energy operators $K$, $K^\pm$ and
$K^\pm_n$, the time evolution of Dirac field operators takes the form (5.39). 

\vskip0.1in
\noindent
{\it Proof}: By (5.22), this is equivalent to show 
$$a(f_\pm(\cdot-t))=e^{iKt}a(f_\pm)e^{-iKt},\eqno(5.46a)$$
$$a^\dagger(f_\pm(\cdot-t))=e^{iKt}a^\dagger(f_\pm)e^{-iKt}.\eqno(5.46b)$$
By (5.23), we have 
$$e^{\pm i\omega t}f_\pm=\int e^{\pm i\omega x'_0}\pi_\pm\gamma^0f(x'_0-t,{\bf x})dx'_0=f_\pm(\cdot-t).\eqno(5.47)$$
To show (5.46), it suffices to prove  
$$a(e^{\pm i\omega t}f_\pm)=e^{iKt}a(f_\pm)e^{-iKt},\eqno(5.48a)$$ 
$$a^\dagger(e^{\pm i\omega
t}f_\pm)=e^{iKt}a^\dagger(f_\pm)e^{-iKt}.\eqno(5.48b)$$
The proof of (5.48) is standard [22]. $\square$

\vskip0.1in
If one replaces the negative energy annihilation operator by a particle
conjugated creation operator, then time evolution is implemented with
positive energy (refer to [11] for details). This completes our work.

\section* {Acknowledgments}
The author would like to thank professor Jonathan Dimock for his
enlightening comments during numerous discussions. The referee's comments
are also appreciated. The author was partially supported by NSF Grant
PHY9722045. 

\section* {References}
$[1]$ Lichnerowicz A 1964 {\it Bull. Soc. Math. France} {\bf 92}
11-100\\
$[2]$ DeWitt B S 1965 {\it Dynamical theory of groups and fields} (New
York: Gordon and Breach)\\ 
$[3]$ Choquet-Bruhat Y 1968 Hyperbolic differential equations on a
manifold {\it Battelle Rencontres} ed B S DeWitt and J A Wheeler (Menlo Park, CA: Benjamin-Cummings)\\ 
$[4]$ Choquet-Bruhat Y, DeWitt-Morette C and Dillard-Bleick M 1977 {\it
Analysis, manifold, and physics} (Amsterdam: North Holland)\\ 
$[5]$ Dimock J 1982 {\it Trans. Amer. Math. Soc.} {\bf 269} 133-147\\ 
$[6]$ Hollands S 1999 The Hadamard condition for Dirac fields and
adiabatic states on Robertson-Walker spacetimes {\it Preprint} gr-qc/9906076\\
$[7]$ Kratzert K 2000 Singularity structure of the two point function
of the free Dirac field on a globally hyperbolic spacetime {\it
  Preprint} math-ph/0003015\\ 
$[8]$ Kay B S 1978 {\it Comm. Math. Phys.} {\bf 62} (1) 55-70\\ 
$[9]$ Furlani E P 1995 {\it Jour. Math. Phys.} {\bf 36} (3)
1063-1079\\ 
$[10]$ Furlani E P 1997 {\it Class. Quamtum Grav.} {\bf 14} (7) 1665-1677\\ 
$[11]$ Jin W M 1999 Dirac quantum fields in curved space-time {\it PhD
Dissertation}\\
$[12]$ Jin W M 1998 {\it Class. Quantum Grav.} {\bf 15} (10) 3163-3175\\
$[13]$ Bjorken J D and Drell S D 1964 {\it Relativistic Quantum Mechanics} 
(New York: McGraw-Hill)\\
$[14]$ Chernoff P 1973 {\it J. Func. Analysis} {\bf 12} 401-414\\
$[15]$ Strichartz R 1983 {\it J. Func. Analysis} {\bf 52} 48-79\\
$[16]$ Reed M and Simon B 1979 {\it Methods of modern mathematical
  physics} Vol. I (New York: Academic)\\
$[17]$ Reed M and Simon B 1979 {\it Methods of modern mathematical physics}
Vol. II (New York: Academic)\\ 
$[18]$ Reed M and Simon B 1979 {\it Methods of modern mathematical
  physics} Vol. III (New York: Academic)\\
$[19]$ Bjorken J D and Drell S D 1965 {\it Relativistic Quantum Fields} 
(New York: McGraw-Hill)\\
$[20]$ Wightman A 1973 {\it Proceedings of Symposia in Pure Mathematics}
Vol. {\bf 23} (D.C. Spencer, ed. AMS, Provincetown)\\
$[21]$ Wightman A 1977 {\it Invariant wave equations} (Velo and Wightman,
eds.) (Berlin; New York: Springer-Verlag)\\
$[22]$ Bratteli O and Robinson D W 1981 {\it Operator Algebras and Quantum
Statistical Mechanics} (New York: Springer-Verlag)

\end{document}